\documentclass{iau}
\usepackage{graphicx}

\title[The Failed Wind in Active Galactic Nuclei]
{The Role of Failed Accretion Disk Winds \\in Active Galactic Nuclei}

\author[Margherita Giustini \& Daniel Proga]
{Margherita Giustini$^1$ \and Daniel Proga$^2$}

\affiliation{$^1$Centro de Astrobiolog\'ia (CSIC-INTA), Departamento de
Astrof\'isica; \\ Camino Bajo del Castillo s/n, Villanueva de la Ca\~nada, E-28692 Madrid,
Spain
 \\ email: {\tt mgiustini@cab.inta-csic.es} \\[\affilskip]
$^2$Department of Physics \& Astronomy, University of Nevada, Las
Vegas;\\ 4505 South Maryland Parkway, NV 89154-4002 Las Vegas, USA     \\email: {\tt dproga@physics.unlv.edu}}

\pubyear{2019}
\volume{356}  %% insert here IAU Symposium No.
\jname{Nuclear Activity in Galaxies Across Cosmic Time}
\editors{M. Povic et al., eds.}

\begin{document}

\maketitle

\begin{abstract}
Both observational and theoretical evidence point at outflows originating
from accretion disks as fundamental ingredients of active galactic nuclei (AGN).
These outflows can have more than one component, for example an unbound supersonic wind
and a failed wind (FW). The latter is a prediction of the simulations of radiation-driven disk outflows
which show that the former is accompanied by an inner failed component,
where the flow struggles to escape from the strong gravitational pull
of the supermassive black hole. This FW component could provide
a physical framework to interpret various phenomenological components of AGN.
Here we briefly discuss a few of them: the broad line region, the X-ray obscurer, and the X-ray corona.

\end{abstract}

\firstsection
\section{Introduction}

The inner structure of luminous active galactic nuclei (AGN) is shaped
by the presence
of winds launched on accretion disk scales, as indicated by their large
terminal velocities
($\upsilon_{\rm{out}}\gg 5000$ km s$^{-1}$ and up to several $0.1$c).
Such winds cannot be sustained by thermal pressure, but must be driven
by either
magnetic or radiative forces (e.g., see for a recent review Giustini \& Proga 2019).
In the case of radiation-driven accretion disk winds (specifically
line-driven, LD),
 a chaotic, dense, struggling inner component is always accompanying the
larger scale successful
 mass outflow: it is the failed wind (FW; Proga, Stone \& Kallman 2000;
Proga \& Kallman 2004; Proga 2005).
\textit{Here ``successful'' or ``failed'' means the ability of inability
of the gas in reaching the escape velocity $\upsilon_{\rm{esc}}$.}
In LD disk wind models, the FW is a fundamental ingredient of the
mass flow of luminous AGN as the wind itself.
In fact, its formation is more robust than the formation of the wind.
Therefore if LD disk wind models hold, observable
quantities related to the FW should be identifiable in AGN as well.
In the following section, we briefly discuss the possible role of the FW in the inner accretion flow of luminous AGN.

\section{Failed line-driven accretion disk winds in the AGN inner structure\label{sec:FW}}

Line driving can deposit much
more momentum in the gas than pure electron scattering,
allowing the launching of material with velocity greater than $\upsilon_{\rm{esc}}$
in sub-Eddington regimes (Castor, Abbott \& Klein 1975).
For LD to be effective, the presence of spectral transitions is
therefore fundamental.
The spectral energy distribution of luminous AGN allows for LD
winds to be launched
from accretion disk scales pushing on the many UV transitions available
(e.g., Murray et al. 1995; Proga, Stone \& Kallman 2000; Proga \& Kallman 2004).
A large X-ray flux is also characteristic of luminous AGN, and while the
X-ray photons will push
on relatively few available X-ray lines, they will mostly concur
to strip the electrons
off the UV-absorbing atoms, thus destroying the many UV spectral
transitions available and effectively
``overionizing'' the wind material (e.g., Dannen et al. 2019, and
references therein).
\textit{The term ``overionization'' is referred to a level of ionization
that is too large to produce the observed UV lines and to sustain LD winds above local $\upsilon_{\rm{esc}}$.}
Therefore the ratio between the UV and X-ray radiation flux is crucial
for the successful launch and acceleration of LD winds in AGN.

In the first models of LD accretion disk winds in AGN, a layer of dense
gas (a \textit{shield}), absorbing the strong ionizing X-ray flux,
was assumed to exist between the X-ray continuum source and the
UV-absorbing wind.
This is the ``hitchhicking gas'' of Murray et al. 1995, which postulated
the presence of this gas just in front of the flow that is
effectively accelerated out of the system. It was speculated that a gradient in pressure would
then cause the hitchhicking gas to accelerate together with the
farther out wind (hence the nickname).

Hydrodynamical simulations performed by Proga and collaborators (Proga,
Stone \& Kallman 2000; Proga \& Kallman 2004)
 showed that for massive (black hole mass $M_{\rm{BH}}>10^8\,M_{\odot}$),
luminous AGN (Eddington ratio $\dot{m} > 0.5$), the accretion flow
settles in a hot polar flow, a fast equatorial outflow,
and an inner transitional zone where the material is lifted up by the
strong radiation pressure.
The gas in the transition zone is exposed to the strong ionizing continuum, loses most or all of
its bond electrons, thus losing line driving force.
This material is unable to reach $\upsilon_{\rm{esc}}$ before getting
overionized, and it falls back toward
the disk.
\textit{The failure or success of the wind is measured in terms of
overcoming or not the local $\upsilon_{\rm{esc}}$}.
The inner FW effectively protects (shields) the material located
farther out from the strong ionizing continuum radiation,
therefore allowing for the successful launch of the wind at radii larger
than where the FW dominates, and where
 the radiation pressure is large enough to overcome BH gravity
(Proga \& Kallman 2004; Risaliti \& Elvis 2010).

In the scenario proposed by Giustini \& Proga (2019), for
$M_{\rm{BH}}\gtrsim 10^8\, M_{\odot}$ and $\dot{m}\gtrsim 0.01$,
the AGN inner structure is dominated by the presence of a LD
disk wind and its inner FW component.
A LD wind is launched at all radii where the local radiation pressure
overcomes gravity; in the inner
portions of disk the wind is unable to escape because of
overionization, and therefore forms a FW.
The AGN appearance will be more or less dominated by the FW (and thus have slower
or faster LD winds), depending on how large is
the X-ray/UV flux ratio as seen by the gas.
Although the Giustini \& Proga 2019 scenario is still qualitative, it
already makes some predictions.
We discuss in the following three phenomenological components of AGN that might be
explained by the FW in AGN, in order of decreasing distances
from the central supermassive black hole (SMBH): the (high-ionization)
broad line region in Section~\ref{sec:BLR}, the X-ray obscurer in Section~\ref{sec:OBS}, and the X-ray corona(e)
in Section~\ref{sec:CORONAE}.

\subsection{The FW as BLR\label{sec:BLR}}

%%% Leighly 2004, NLS1; Gallagher 2005, BAL QSOs; alpha_ox and

The broad line region (BLR) is a fundamental ingredient of luminous AGN: it
consists of gas photoionized by the AGN continuum and whose motion responds to
the gravitational potential of the central SMBH (e.g., Peterson et al. 2004).
%The BLR gas imprints emission and absorption lines in the AGN spectra; the study
%of emission and absorption line profiles allows to get insights into the
%geometry and the dynamics of the BLR flow.
The BLR is phenomenologically divided into a low-ionization (e.g.,
Mg II, H$\beta$) and a high-ionization (e.g., C IV, Ly$\alpha$) component
which show distinct kinematics (e.g., Marziani et al. 1996).
In particular, a difference in peak position between low-ionization and high-ionization emission lines
is indicative of strong radial motions of the gas producing the latter (Gaskell 1982).

The high-ionization emission lines in luminous AGN can in fact be blueshifted by several hundreds
(up to thousands) km s$^{-1}$ with respect to the low-ionization emission lines
and the host galaxy (e.g., Richards et al. 2011): part of the high-ionization BLR must be a wind.
Most of the low-ionization BLR likely originates at large scales, close to the dust sublimation radius,
 where radiation pressure on dust grains can form an outer wind and a FW (Czerny \& Hryniewicz 2011).
 Radiation pressure on UV lines, that forms a LD accretion disk wind
 and an inner FW, can explain instead the bulk of the high-ionization emission lines
 (Proga \& Kallman 2004).
 Part of the low-ionization emission lines can also be produced
 within dense clumps at the base of the wind, and would then also be blueshifted (Waters et al. 2016).

Blueshifted broad absorption lines in high-ionization UV transitions\footnote{
Low-ionization broad absorption lines are observed in a small fraction (about 5-10\%) of
 broad absorption line quasars, the low-ionization BAL QSOs; those who do not display them
 are called high-ionization broad absorption line quasars.}
 are also observed in a large
number of luminous AGN (up to 40\%, Allen et al. 2011).
These are the so-called broad absorption line quasars (BAL QSOs), and display the most direct evidence for
the presence along the line of sight of strong winds, which can reach velocities of several $0.1$c.
Such high velocities must be produced close to the SMBH, on accretion disk scales.
The presence of the broad absorption troughs alone indicates that a lot of momentum has been deposited
in the gas by radiation.
Remarkably, the most recent observations of large samples of AGN have demonstrated that high-ionization
BAL QSOs correspond to quasars in general (Rankine et al. 2020), thus
supporting accretion disk wind scenarios for luminous AGN in general.
If driven by radiation, a disk wind will be accompanied by the inner FW,
and this will also contribute to the emission of the BLR. But how?

As summarised in Giustini \& Proga 2019, in LD disk winds scenarios
the BLR appears in luminous AGN at $\dot{m}\gtrsim 0.01$,
when the inner accretion and ejection flow consists of a disk, LD wind, FW,
and inner X-ray source; the LD wind + FW then produce the high-ionization BLR.
In particular, the production of the symmetric portion of the high-ionization BLR
is associated to the FW, while its blueshifted and blue-skewed portion,
to the wind itself.

The strongest (fastest, densest) LD disk winds are those produced in AGN with a low
X-ray/UV flux ratio (\textit{X-ray weak}), either because of high $\dot{m}$ and/or
a large $M_{\rm{BH}}$.
These have a vast radial zone of the inner flow dominated by winds,
and only a small inner region where the wind fails.
Thus they produce winds with a large range of velocities,
including large terminal velocities $\gg 5,000$ km s$^{-1}$ and up to several $0.1$c
when launched in the innermost regions of the disk.
On the contrary, in the case of AGN with a large X-ray/UV flux ratio (\textit{X-ray bright}),
a larger inner region of the accretion flow is dominated by the FW.
In these AGN, successful winds are only launched at larger scales,
thus reaching lower terminal velocities.

In LD disk winds scenarios, the BLR of X-ray weak AGN is dynamically dominated by the wind:
the emission lines of e.g. C~IV are strongly blueshifted and blue-skewed.
Their equivalent width is lower than the one of the same emission lines
produced in X-ray bright AGN: here the dynamics of the BLR is dominated
by the FW, that does not reach $\upsilon_{\rm{esc}}$. The emission lines have a larger equivalent
 width, a more symmetric profile, and little or no blueshift with respect to the
 redshift of the host galaxy.
In other words, the FW extent regulates the extent of the symmetric, non-shifted BLR at the
expense of the skewed, blueshifted BLR produced in the wind.
When the disk is observed through the wind, X-ray weak AGN will
display deeper, broader, and more blueshifted absorption troughs compared to X-ray bright AGN.
Broadly speaking, the first type of AGN would correspond to the population A of quasars along their main sequence,
while the second type of AGN to their population B (Sulentic et al. 2000; Sulentic \& Marziani
2015).

\subsection{The FW as the obscurer}\label{sec:OBS}

The presence of dense, variable layers of X-ray absorbing gas on
BLR-scales has been recently inferred by high-quality
observations of local Seyfert 1 galaxies (Kaastra et al 2014; Ebrero et
al. 2016; Mehdipour et al. 2017; Kriss et al. 2019).
This gas is called ``obscurer'', as it absorbs the X-ray continuum flux
and obscures the view of the strong ionizing X-ray continuum to the
material located further out. This further out material, in fact, responds
to the changes in X-ray ionizing flux, as strong UV absorption lines are
observed emerging in concomitance with the appearance of strong X-ray
absorption.

In LD accretion disk wind scenarios, the FW is the material
located close to the source of X-rays,
that gets all the ionizing continuum, and thus fails reaching  $\upsilon_{\rm{esc}}$
 and falls back toward the disk.
The FW motion is complex: highly dynamical, with locally variable
motion made of upward and downward components,
and dense filaments and knots embedded in a much hotter medium (Proga 2005).
The FW absorbs the X-ray continuum photons, effectively shielding the
gas located farther out that can be then accelerated by radiation
pressure on UV spectral lines.
The FW has therefore all the characteristics to be identified with the
``obscurer'' of local Seyfert galaxies.

\subsection{The FW as the X-ray warm corona(e)}\label{sec:CORONAE}
Much closer to the central SMBH than the wind, but maybe partially
co-spatial with the inner FW, lies the source of X-ray photons.
The X-ray radiation is the clearest signature of accreting BHs, yet its
physical origin is still unclear.
We know that some compact and hot region must be responsible for the
bulk of the intense and variable X-ray emission of AGN,
and we call it X-ray ``hot corona''.
The X-ray hot corona has become a synonym for a low-density (optical depth
$\tau \ll 1$) medium full of hot (temperature $kT_{\rm{e}}>100$ keV)
electrons, that by interacting with the much slower UV photons emerging
from the thermalized accretion disk, increase their energy through
inverse Compton scattering (e.g., Haardt \& Maraschi 1993).
This hot coronal emission is able to overionize the wind material,
therefore concurring in destroying the wind,
i.e., creating a FW.
In recent years, the presence in the inner regions of luminous AGN of
material also able to Compton-upscatter UV photons, but with much lower
temperature and much higher optical depth ($kT_{\rm{e}}\sim 100-300$ eV,
$\tau\sim 10$), has emerged: this is called the X-ray ``warm corona''
(Done et al. 2012; Mehdipour et al. 2015; Petrucci et al. 2018).
These general physical properties of the X-ray warm corona
look similar to those of the FW.
An exchange of energy is expected between the source of hard X-ray photons (the hot corona)
 and the FW, with dense knots within the FW able to emit X-ray
bremsstrahlung and thus, in the most extreme cases, switching the main
cooling mechanism for the X-ray emitting plasma (Proga 2005).
In these cases the warm corona can dominate over the
hot corona in terms of density and hence matter cooling/radiation emission,
and produce genuinely hard X-ray weak AGN where most of the
flux of photons at $E\gtrsim 2$ keV is suppressed: the FW would then
become the X-ray corona itself.

\section{Conclusions}

Accretion disk winds have been recognized as fundamental ingredients of the inner regions of luminous AGN.
In the case of LD disk winds, the inner FW component might help interpreting in a physical framework
phenomenological features of AGN such as the high-ionization BLR, the obscurer, and
the X-ray coronae. The FW solutions of the inner accretion and ejection flow of AGN deserve
further attention, in order to assess whether they can change significantly the physical
and geometrical structure of the very inner accretion
flow around highly accreting SMBHs.

 \section{Acknowledgements}

 MG warmly thanks the IAU 356 symposium organizers for a memorable,
transformational meeting.
We thank G. Richards, G. Miniutti, E. Lusso, and M. Mehdipour
 for interesting discussions.
MG is supported by the ``Programa de Atracci\'on de Talento'' of the
Comunidad de Madrid, grant number 2018-T1/TIC-11733, 
and by the Spanish State Research Agency  (AEI) Projects number
ESP2017-86582-C4-1-R and ESP2015-65597-C4-1-R.
This research has been partially funded by the AEI Project number
MDM-2017-0737 Unidad de Excelencia ``Mar\'ia de Maeztu'' - Centro de
Astrobiolog\'ia (INTA-CSIC).

\end{document}